\documentclass{svproc}
\usepackage{url}
\usepackage{graphicx}
\usepackage{caption}
\begin{document}
\mainmatter              
\title{Ethical Analysis on the Application of Neurotechnology for Human Augmentation in Physicians and Surgeons}
\titlerunning{Ethical Analysis}  
%
\author{Soaad Q. Hossain\inst{1} \and Syed Ishtiaque Ahmed\inst{2}}
\authorrunning{Hossain and Ahmed} 
%
\tocauthor{Soaad Q. Hossain and Syed Ishtiaque Ahmed}
\institute{Department of Computer and Mathematical Sciences, Department of Philosophy, University of Toronto Scarborough, Canada\\
\email{soaad.hossain@mail.utoronto.ca}
\and
Department of Computer Science, University of Toronto, Canada\\
\email{ishtiaque@cs.toronto.edu}
}

\maketitle              

\begin{abstract}
With the shortage of physicians and surgeons and increase in demand worldwide due to situations such as the COVID-19 pandemic, there is a growing interest in finding solutions to help address the problem. A solution to this problem would be to use neurotechnology to provide them augmented cognition, senses and action for optimal diagnosis and treatment. Consequently, doing so can negatively impact them and others. We argue that applying neurotechnology for human enhancement in physicians and surgeons can cause injustices, and harm to them and patients. In this paper, we will first describe the augmentations and neurotechnologies that can be used to achieve the relevant augmentations for physicians and surgeons. We will then review selected ethical concerns discussed within literature, discuss the neuroengineering behind using neurotechnology for augmentation purposes, then conclude with an analysis on outcomes and ethical issues of implementing human augmentation via neurotechnology in medical and surgical practice.
\keywords{neurotechnology, ethics, augmentation, enhancement, brain-computer interface, physicians, surgeons, patients, harm, global, social, justice, malicious brain-hacking, personhood, brain data, discrimination, rights, medical practice, cognition, senses, action}
\end{abstract}
\section{Introduction}
The demand for medical and healthcare professionals continue to rise as the overall world population continues to increase. This demand is further increased when pandemics such as the SARS-CoV-2 outbreak, also known as the coronavirus disease COVID-19, occurs. The problem is that there is and may always be a shortage of especially medical practitioners due to challenges and barriers that exist to make it difficult for individuals to become physicians or surgeons. Even if there were sufficient medical practitioners, the other problem is the lack of resources that hospitals and institutions alike possess. Given that resources are already a problem for hospitals, having sufficient clinicians would only make that worse. With recent advances in neuroscience and neurotechnology paving ways to innovative applications that cognitively augment and enhance humans in a variety of contexts, a solution to this would be to use neurotechnology to augment the cognition and senses of medical practitioners to then augment their actions [13]. In doing so, this will allow them to make optimal decisions and actions in less time, enabling them to treat more patients. As such, this makes the solution along with the research and discussions pertaining to it highly important as not only does the solution impact directly impact physicians and surgeons, but it also directly impacts patients as well. Additionally, as with any technological intervention, it is important that not only are the technical aspects of the intervention are analyzed and discussed, but the ethical aspects are as well. In approaching this technological intervention, we performed an ethical analysis on the application of neurotechnology for human augmentation in physicians and surgeons, and found that even in the best-case scenario, applying neurotechnology for human augmentation in physicians and surgeons has a significant negative impact on individuals, communities and countries. 
\\ \\ 
     In this paper, we argue that applying neurotechnology for human augmentation to augment physicians and surgeons, and can cause personal identity, discrimination and financial issues for physicians and surgeons, and lead to patients being harmed. The way that the paper is structured is as followed: we first describe the augmentations and neurotechnologies that can be used to achieve the relevant augmentations for physicians and surgeons, then review selected ethical concerns discussed within literature particularly focusing on human rights, human-computer interaction, data, brain-computer interface, global bioethics and drug development, and discuss the neuroengineering behind using neurotechnology for augmentation purposes. We then conclude with an analysis on outcomes and ethical issues of implementing human augmentation via neurotechnology in medical and surgical practice. The ethical analysis specifically focuses on social issues, global health inequality and health migration, and patient harm, and includes an assessment on personhood with respect to the neurotechnology users (i.e. the physicians and surgeons). In this paper, we assume that all neurotechnologies mentioned always succeeds in providing a person with augmentations, but the type of augmentation provided by a neurotechnology is based on what its capable of. The motivation behind this assumption is to allow the paper to address a possible best-case scenario with neurotechnology and human augmentation in physicians and surgeons, and focus on ethical issues that arise with this scenario. 
\section{Human Augmentation and Neurotechnology}
\subsection{Types of Augmentations}
     Human augmentation can be formally defined as an interdisciplinary field that addresses methods, technologies and their application for enhancing cognitive abilities, senses and actions of humans [32]. In enhancing these leads us to augmented cognition, augmented senses, and augmented actions. Augmented  senses focus on enhancing one’s ability to see, hear, smell, taste and touch. Augmented cognition focuses on enhancing one’s memory (short-term and long-term), attention, learning capabilities, awareness, and knowledge. Augmented action can be broken down to two parts: motion augmentation and gesture augmentation. Motion augmentation would simply be improving one’s ability to move, enabling them to take actions that they may not normally be able to do such as carry very heavy objects or run at a faster speed. Gesture augmentation is similar to motion augmentation, except that it focuses more movement and positioning of one’s hand and head. For instance, being able to keep your hand in a specific position for a long amount of time without fatigue and shaking would be considered augmented gesture and not motion augmentation. The end goal of both augmentations is that the augmentation should allow the person to perform an action optimally. From all the augmentations mentioned, we will only focus on augmented senses (specifically augmented vision and augmented touch), augmented cognition, and augmented action (specifically augmented gesture). The motivation behind these augmentations in particular is because these augmentations best allow physicians and surgeons to better diagnose and treat diseases and disorders. 
\subsection{Augmentations Through Neurotechnology}
     From current technologies that are being developed, one such technology that can be used to obtain these augmentations is neurotechnology. Neurotechnology can be formally defined as an interdisciplinary field that combines neuroscience, engineering and technology to create technical devices that are interfaces with human nervous system. Within literature on neurotechnology, there has been a considerable amount of work carried out on neurotechnologies for cognitive enhancement, specifically focusing on brain-computer interface (BCI), also known as neuroelectronic interface [2], applications. Based on previous studies, neurostimulation techniques, such as transcranial electric stimulation (tES) and transcranial magnetic stimulation (TMS), can be used to improve performance in different cognitive domains; these cognitive domains include perception, learning and memory, attention, and decision-making [13]. Accordingly, this can be used to accomplish augmented vision and part of augmented cognition. Neuromodulation techniques, such as transcranial direct current stimulation (tDCS), can also be used for memory, learning and attention [13]. Specifically, for tDCS, studies have shown that it could improve performance in verbal problem-solving task and other areas within complex problem solving [13]. This allows for further enhancement of specific cognitive domains, as desired. 
\\ \\ 
     To address augmented touch while also addressing augmented action, a combination of tactile sensors and neuroprosthetics. Tactile sensors allow for perceiving the location of the point of contact, discerns surface properties, and detection for eventual slippage phenomena during the grasping and handling of an object [11]. Neuroprosthetics are devices that can enhance the input and output of a neural system [5]. Combined together, they can allow for augmented gesture through enhancing the person’s dexterous interaction with objects by ensuring that there is a correct application of forces between the contact surfaces of the person’s hand and fingers and those of the object, and having the neuroprosthetic distinguish brain activity patterns corresponding to the intention to act from control signals of the hand itself [38]. To summarize, the neurotechnology that can be used to accomplish the augmentations of interest is neurostimulation, neuromodulation, tactile sensors and neuroprosthetics. How each of these neurotechnologies accomplish augmentations will be discussed later in the paper when We discuss how neurotechnology can be applied in medical practice.  
\section{Ethics in Neurotechnology and Human Enhancement}
\subsection{Human Rights in Neurotechnology}
     While neurotechnology has the can help achieve human augmentation, it also has the potential to impact human rights [22]. Three of the areas which neurotechnology is said to impact is the right to mental integrity, freedom of thought, and freedom from discrimination. The right to mental integrity is a concern with neurotechnology due to the problem of malicious brain-hacking. Malicious brain-hacking is defined as the neuro-criminal activities that directly influence neural computation in the users of neurotechnological devices in a manner that resembles how computers are hacked in computer crimes [22]. For BCI applications, malicious brain-hacking can take place within BCI through having the malicious agent (e.g. hacker) attack the level of measurement or decoding and feedback of the BCI, or through the malicious agent manipulating the person’s neural computation through the BCI application. In such cases, the malicious agent can add noise or override the signals sent to the neurotechnological device with the objective of reducing or removing the control of the user over the BCI application [22]. Furthermore, the malicious agent can also hijack the BCI user’s voluntary control. For instance, a malicious agent can override the signals sent from a BCI user, then hijack a BCI-controlled device (e.g. smartphone) without the BCI user’s consent [22]. Ultimately, what results from malicious brain-hacking is both the user’s mental privacy and their brain data being at risk due from the loss of right to mental integrity. The loss of right to mental integrity leads to the loss of freedom of thought as when malicious brain-hacking occurs, the BCI user is no longer able to freely think of things due to the restraint placed by the malicious agent. 
\\ \\
One other human rights issue with neurotechnology is one that overlaps with a health rights issue from human enhancement. The ethical issue from human enhancement pertains to how the ability to augment one’s physical or mental performance raises several issues about fairness and justice regarding how augmenting technologies, such as neurotechnologies, should be accessed or regulated [20]. What follows from this is the question of whether they are intended for mass consumption or restricted to humans with identifiable impairments and disabilities. In the case where neurotechnologies are intended for mass consumption and becomes commercialized, as with any other technology, it will likely be expensive at the beginning [20]. Consequently, this could generate and even exacerbate societal divisions among the population or among inhabitants of different countries [20]. Those of a low socio-economic status would not be able to afford to purchase neurotechnologies. Subsequently, a digital divide is created between those that can afford neurotechnology, and those that cannot. This leads to health right issue as the digital divide creates an injustice to those of low socio-economic status that cannot afford neurotechnology; they are indirectly being deprived of something that is needed for their health and wellbeing. With the relationship between poor mental health and the experience of poverty and deprivation being well studied and an association between the two factors being established, communities deprived of neurotechnology can cause those communities to experience higher levels of deprivation and unemployment rates, leading to them experiencing higher rates of psychiatric admissions for psychotic as well as non-psychotic conditions and suicidal behavior [27]. 
\subsection{Discrimination from Augmentation}
     The ongoing human rights concern regarding  neurotechnology an freedom from discrimination is that the use of neurotechnology will lead to discrimination among individuals and groups. This concern is not only one that is valid, but also strong evidence to support it. Throughout history, humans have shown that existing moral code is often broken in practice, and that observable differences between people, such as differences in race, gender, ethnicity, religion, sexual orientation and/or ability, tend to lead to moral and social inequalities [7]. Given that there will be a clear difference between those that adopt augmentation technologies and those that do not and the actions that humans have taken in the past and continue to take to this day, it is very likely that human augmentation will lead to new, unjustified inequalities, and may even undermined the core notion of moral equality used in Western societies [7].  Expanding on the unjustified inequality, there is a growing concern regarding how new human augmentation technologies can weigh more than personal experiences and benefits [32]. Due to the cognitive enhancement that augmentation technologies provide, those that utilize them will have a stronger say on matters than those that rely solely on their personal experiences and benefits. Also, there is a concern that neglecting potential negative effects of such technologies to daily life and society can create problematic scenarios for the future [32]. One such potential negative side effect comes from the pressure in adopting augmentation technologies. For a long time and even now, people frequently experience prejudice if their bodies or brain function differently from those around them [40]. With the integration of augmentation technologies such as neurotechnology within society, this will lead to people feeling pressure to adopt those technologies, and is likely to lead to change societal norms. In turn, this will lead to issues of equitable access and generate new forms of discrimination [40]. 
\\ \\ 
     To describe the forms of discrimination, we refer to self-identity, industry practice, and history. In self-identity, moral and social importance plays a crucial role in how one people feel about themselves, and it plays a strong role in determining people’s intentions, attitudes and behaviors [7]. Consequently, a poorly developed self-concept could bring about low self-esteem, level of confidence, purpose and reason to live, and level of motivation. With groups and communities determining collectively how one is treated by others, poorly formed self-concept could make one subject to discrimination. Within industries, the discrimination can take place as earlier as the job application process to later while the person is employed with the company. In industry practice, it is common for companies to hire the person that they think are best for the position. With neurotechnology providing its users with augmentations, a complaint regarding this is that this provides those people with an unfair advantage that will allow them to secure jobs more easily, and that those that are augmented may view those that are not augmented as inferior. This unfair advantage can lead to discrimination within industries through employment. 
\\ \\ 
     The unfair advantage from augmentations can have serious implications in the workplace especially once employers, administrators and decision-makers realize that those with augmentation perform better than those without them [10]. The realization will likely lead to them having a preference for those with augmentations and a bias against those that do not have any augmentations, leading to things such as them inquiring beforehand (or on the job for current employees) whether an applicant has undergone an augmentation or utilizes neurotechnology, and making decisions simply based on this information along with selected other pieces of information from the applicant (or employee). This can lead to prejudice, resulting in non-augmented people being discriminated against, creating justice and equity issues. Note that this type of discrimination is not limited to employment settings, and extends as far as in areas such as families and academic institutions. Between children, parents will give preference to the child that performs best academically and professionally. Similarly, with academic institutions, the institutions will prefer and reward those that perform best academically, and treat those that perform best academically better than those that perform worse regardless of whether they have undergone any sort of augmentation or not. We know this as in many academic institutions, if students do not perform well enough academically, they are either removed from their program, suspended or both, and those that perform well academically are rewarded awards, research and other opportunities, and scholarships. As a result, in the long-term and possibly even in the short-term, the implications from the unfair advantage from augmentations can lead to a collapse in the social process of education, care, interactions, relationships and more as impartiality, open-mindedness, nondiscrimination, acceptance, and unbiasedness would be minimized or lost completely in some groups and societies.  
\\ \\ 
     With neurotechnology and human enhancement possibly being enjoyed by the already privileged, if this possibility is realized, then this will lead to a failure of distributive justice [18]. Of all communities, marginalized communities will suffer the most from this failure. A major challenge experienced by underprivileged communities comes from the fact that not only are they are often excluded from opportunities, health services, development programs, resources and more, but they often struggle to raise their voice in public discussions due to the lack of access and autonomy [4]. As a result, underprivileged communities, including women, LGBTQ groups, refugees, people with physical or mental disabilities, people of low socio-economic status, and those that are less educated are often ignored, deprived, or discriminated [4]. Among them, marginalized groups in those communities are even more deprived, which this deprivation are sometimes due to their physical or economic vulnerabilities, or various social and cultural practices [4]. For example, in multiple places in the Indian subcontinent, women are not allowed to use mobile phones due to social and cultural reasons [4]. Consequently, this deprivation prevents them from obtaining the benefits from it -  one of them being that it can be used to cope with everyday stress [28]. In the case where the group of Indian women were members of the LGBTQ group, this deprivation of technological device can further negatively impact their mental health. LGBTQ people have and continue to experience various forms of oppression and discrimination worldwide [21]. The oppression is in the form of harassment and violence while the discrimination is experienced in areas such as employment, housing, access to education and human services, and the law. This oppressions and discrimination along with rejection, violence and harassment have been shown to have negative physical and mental health effects on both LGBTQ people [21]. Combining the primary societal stressors that many LGBT people experience, that lead to “minority stress” [21], with deprivation of mobile phones leads to those Indian women having less effective coping mechanism options which are needed to deal with both everyday stress and minority stress. Consequently, this results to there being major mental health disparities between those Indian women within the LGBTQ population and other individuals. From just the transgender individuals alone, the mental health disparities can be severe enough to the point where the difference is that those individuals have higher odds of depression symptoms and attempted suicides [39]. What the failure of distributive justice will likely do is aggravate the everyday stress and minority stress experienced by this marginalized group and others to an extent where the majority of of the marginalized population will show depression symptoms, and a substantial amount of them may even commit suicide. As such, the already marginalized communities will suffer more than they already are. The repercussions of this will have negative impacts on both a local and global level. 
\subsection{Autonomy and Brain Data}
     Neurotechnology not only has the potential to impact human rights, but is can also impact autonomy, confidentiality and protection. Note that it is possible to obtain information through BCIs that can be extracted to reveal security breaches [26]. Neurotechnology such as BCI relies on multiple types of probabilistic inference to be operationalized, for one to use  any sort of probabilistic inference requires a large amount of data [19]. This data, consisting of neural information, is of high value to companies and individuals as neurotechnology produces raw data in a way that enables a more direct detection pathway of the neural correlates of mental processes, such as interests, intentions, mood and preferences [23]. Additionally, the neural data includes rich and personally identifiable sources of information that could be aggregated by data handlers to apprehend or predict parts of health status, preferences and behavior [23]. An example of where neural information would be very useful is marketing and advertising. The field of marketing has been interested in BCI research as they are interested in utilizing a neuromarketing-based approach to marketing and advertising [3]. Algorithms that are designed and used to target advertising can better calculate things like insurance premiums through using neural information [40]. Accordingly, investment in neural data (and computer scientists) is something that marketing firms are striving to acquire, heavily investing time, effort and money to obtain it.   
\\ \\
     Given the value of neural information, these motivate malicious agents, companies and corporations to perform malicious brain-hacking, or retrieve, aggregate disseminate and use the information of the neurotechnology users without their informed consent. In the case of malicious brain-hacking, after a malicious agent has successfully hacked a device, rather than changing signals or taking control, they can instead extract the data from the device, then abuse to them to as they desire. Similarly, for companies and corporations behind a neurotechnological application, they can either take the data collected from the neurotechnology users as hostage of the users or sell the data to another entity or individual, with the purpose of making more money. This has several consequences, including loss of ownership of one’s own data, loss of privacy, increase in level of distrust, data used against them (e.g. for insurance purposes),  Furthermore, with the malicious hacker or entities not getting the consent from the users to take and sell their data, the autonomy of the users is not respected.
\subsection{Global Bioethics and Drug Development}
     The aftermath from a situation of malicious brain-hacking, neural data hostage, or neural data sharing does not just impact an individual, but it impacts society altogether. However, of the three, We will focus on neural data sharing, continuing the discussion from the point where what can occur after an individual, corporation, or organization has obtained neural data as this is one of the biggest concerns expressed within literature due to the impact that neural data sharing has on society at large. What a person or entity will do with the neural data is difficult to determine. However, it is easier for us to get a sense on what some individuals and companies will do with the data compared to others. With the advances in artificial intelligence and machine learning, when combining learning algorithms with brain data obtained through BCI or other neurotechnologies, they can lead to fruitful results for things such as drug development. However, using BCI for drug development is not a simple task. Contemporary drug discovery strategies rely on the identification of molecular targets associated with a specific pathway and subsequently probe the role of a given gene product in disease progression [6]. Subsequent to the identification of a relevant bioactive compound, further compound screening can be achieved through phenotypic screens. With phenotypic screens, they are capable of capturing complex cellular level behaviors that are physiologically relevant and central to a pathology without relying on the identification of a specific drug target or a corresponding hypothesis concerning its pathological role [6]. For that reason, phenotypic screens are gradually becoming more utilized in drug discovery as using a phenotypic approach can generate novel treatments for diseases and disorders which have either complex or unknown mechanisms [6]. When using a phenotypic approach onto a nervous system, it provides a way to find compounds and genes active in a nervous system without assumptions about relevant molecular targets, which can possibly lead to new therapeutic targets, new disease biology, and new compounds or genes [14]. The challenge is that phenotypic screens require a quantifiable phenotypic alteration, such as neuronal survival or changes in fluorescently tagged protein expression in order to generate a usable output [14]. Conveniently, BCIs can provide brain data containing neuronal survival [2]. With BCIs being able to extract quantifiable phenotypic alteration and classification algorithms being able to understand BCI’s messages, we can conclude that pharmaceutical companies that purchased neural data will most likely use it with a phenotypic approach to discover, develop and patent new generic drugs. The problems that follow this are societal and global health issues. 
\\ \\
     In the past, pharmaceutical companies have consistently argued that high prices and multilateral patent protection afforded by Trade-Related Aspects of Intellectual Property Rights are reasonable rewards for highly expensive and often fruitless research and development of pharmaceutical products, which are ultimately of benefit to society [9]. Furthermore, with pharmaceutical research and development costs being substantial and given the low success rates, pharmaceutical companies argued that profits from the few successful products are essential [9]. Consequently, the problem with pharmaceutical companies researching and developing drugs using neural data and phenotypic approach is that it will likely be very expensive as not only will the companies have to pay a price for conducting and developing the drug, but they also have to pay a price for neural data based on the price set by the third party. For them to generate the data can be even more costly for them as on top of cost associated with conducting data collection, hardware costs from neurotechnologies and costs associated with data preprocessing and augmentation will need to be included in the total cost. Furthermore, with patented drugs preventing others from making or using the generic drug without obtaining a license from the patent holder, most pharmaceutical companies refuse to grant licenses so they can benefit from having a limited monopoly on the drug [33]. As a result, not only will they likely enforce their patents on those generic drugs, but there is a chance that they will also charge a hefty amount for those interested in purchasing the license for it for therapeutic purposes. In turn, this will lead to affordability, accessibility and health issues experienced by the portion of the population unable to afford the drugs. On a global level, with the pharmaceutical companies enforcing their patents, this will prevent developing countries from manufacturing the generic versions of those drugs, creating affordability, accessibility and health issues experienced by those countries [8]. 
\\ \\
     While it was easy to attempt to determine how pharmaceutical companies will neural data, in the case of the military, for example, is not as obvious. The US Department of Defense, for instance, is currently exploring the use of neurotechnologies and neural data [26]. It is said that in pursuing neurotechnology, they are using it with the goal of restoring function following trauma from war and to develop programs involving improving human training and performance [26]. From a first glance, it seems that they are doing it to help better those within the military recover or become a better version of themselves. However, from a closer look, this can be seen as the military trying to get an advantage over other countries [26]. To make neurotechnologies’ use more extreme, yet more effective for the military and even politicians and corporations such as those that sell firearms, exploration of neurotechnologies such as brain implants by the military can be used in a way that create a potential for terrorism [32], leading to beneficence issues. This is an issue as this indirectly promotes warfare between countries, which the consequences from any sort of warfare which neurotechnology contributed to includes an increase in casualties, environmental damage, loss of resources and infrastructure, psychological and neurological disorder cases, loss of talent, and more. These casualties and other losses are a major injustice to the innocent civilians caught in the middle of the warfare [31]. 
\section{Augmentation of Physicians and Surgeons}
\subsection{Neuroengineering for Augmentations}
     Before moving onto how neurotechnology can be applied to medical practice, it is important to have a foundational understanding of how the neurotechnologies that were mentioned earlier accomplish their respective augmentations. How neurostimulation techniques accomplishes enhancing perception, learning and memory, attention, and decision-making within a person is by using invasive or non-invasive electrical stimulation systems on the person [17], to then apply electricity to affect the person’s central nervous system [37]. Similarly, how neuromodulation accomplishes enhancing memory, learning, attention and complex problem solving within a person is by sending electrical or pharmaceutical agents to a target area of the person’s brain to alter nerve activities [24]. While there are several types of tactile sensors, the tactile sensor that would be best for achieving augmented touch and augmented action is a tactile sensor using CNT-based nanocomposites because not only are they mechanically flexible, robust and chemically resistant needed to achieve the flexibility of human skin, but they also require lower conductive phase concentration, which leads to better mechanical properties of the composite [12]. The tactile sensors would be placed onto a sensor glove, then worn by the user.
\\ \\
     The way they work in achieving augmented touch is through optimizing the use volume changes in sensitive materials to detect pressure [12], in a way that allows the tactile sensors to be capable of [11]: (1) evaluating mechanical compliance and surface texture properties of objects; (2) sensing an applied force by evaluating the magnitude and direction; (3) spatially discriminating the point of application of a force; and (4) Emulating the dynamic behavior of human mechanoreceptors while tracking tactile stimuli overtime. Furthermore, the design criteria for the tactile sensors must consider spatial resolution and range of applied force in order to properly achieve augmented touch [12]. Accordingly, neuroprosthetics accomplishes augmented gesture and augmented touch by utilizing tactile sensors with the BCI to receive relevant sensory information concurrently with letting the user directly control the output behavior of the prosthesis (e.g. artificial arm movement); in doing so, this recreates the control-feedback loop [1]. The signals produced by the loop provide output for controlled muscle contractions, input and feedback from sensory organs (e.g. position), creating a bidirectional pathway through which we explore and manipulate our environment [1], enabling for augmented gesture and augmented touch to take place. Figure 1 illustrates the process of augmented gesture. The approach used in the figure incorporates the neurofeedback approach – an approach that is promising for enhancing the performance of the brain [3]. 

\begin{figure}
  \includegraphics[width=\linewidth]{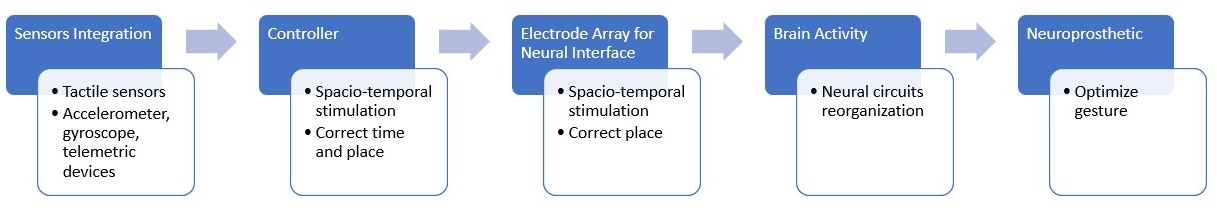}
  \caption{Diagram displaying the steps taken to obtain augmented gesture. First, the tactile sensors that are places on the hands or fingers of the surgeon. Once measurements are obtained, then they are sent to a controller (e.g. BCI) that evaluates, refines and translates the measurements into stimulations that neurons can react to. Upon receiving the stimulations, the neural circuits within	the neurons reorganize themselves in a way that allows the surgeon’s brain to update itself on how to better perform the gesture. The gesture is then realized through neuroprosthetics combined with  either a prosthetic or the surgeon’s hand. 
}
  \label{fig:figure1}
\end{figure}
\subsection{Augmentations in Surgery}
     Knowing how neurotechnology can realize augmentations within humans enables us to address how such technology and augmentations will impact medical practice. For this section, we will specifically focus on the practice of surgery. In this case, the users of the neurotechnology are surgeons. Within surgical settings, augmented vision, augmented touch and augmented gesture will enable surgeons to perform surgeries with a higher level of precision and accuracy by allowing them to see details more clearly, utilize surgical tools and methods better, and perform surgical procedures and operations better. As a result, the success rate of surgeries performed by surgeons utilizing neurotechnology will increase, which is good for multiple reasons; the main reason being that successful surgery will enable the patient to live longer and healthier life, which is helpful for the individual, their family and society. Augmented cognition for surgeons enables surgeons to be able to learn new surgical procedures that are less invasive, shorten operative times, lessens costs and reduce the likelihood of complications. Furthermore, it will allow surgeons to be more knowledgeable in surgical techniques and competent in the ability to recognize the limits of their professional competence [36]. The benefits of these can best be understood through a bioethics perspective, with specifically using the principle of beneficence, non-maleficence and justice, In terms of beneficence, in neurotechnology and the augmentations empowering surgeons, they enable surgeons to take actions, make suitable surgical judgements to assess the risks, burdens and benefits, and approach surgeries in a way that that respects and is in the best the interest of their patients. In addition, their improved cognitive abilities will allow them to stay up to date with activities and update their knowledgebase as required. 
\\ \\
     With having surgeons keeping themselves professionally and educationally advanced, this will help ensure that they will provide the highest standards of patient care and the lowest rates of complications [36]. This leads to non-maleficence as higher standards of patient care is accompanied with better communication and improvement in the type of care being offered (e.g. palliative care),  and lower rates of complications leads to a decrease in the likelihood of a medical error occurring and minimization of harm caused to the patient. Lastly, with the surgeons being able to perform tasks better, this will decrease the time needed for surgeries, and can decrease the time of the patient’s stay in the hospital. From a justice standpoint, this is crucial as just as much as health burden is an issue for patients, so is economic burden. With surgeons successfully performing surgeries on their patients, patients will not need to stay in the hospital for long, enabling them to be able to live back their normal life and do their regular activities, such as working and spending time with their family. 
\subsection{Augmentations in Medical Practice}
    For this section, we will focus  on medical practice that do not involve surgery. In this case, the users of the neurotechnology would be physicians. Within non-surgical settings, augmented vision and augmented cognition will allow physicians to better perform diagnoses. For instance, given that the physician will have enhanced vision and cognitive abilities, analysis of scans from diagnostic radiology exams such as computed tomography (CT) and fluoroscopy would be better performed as physician will be able to see details more clearly and better spot indicators  on the scans. As such, this will allow physicians to make more accurate diagnoses, which can result in better understanding assessment, planning and execution of treatment, communication between the physician and the patient, maintain of the physicians’ fidelity and responsibilities, and care for the patient. These help with achieving non-maleficence as effective diagnosis and communication significantly reduces the amount of harm within patients on both a physical and mental level. Good diagnosis helps with increasing the chances that a correct treatment will be used, and prevents moral distress in patients and their family. Additionally, and most importantly, with neurotechnology, augmented vision and augmented cognition enhancing physicians’ diagnoses performance, this leads to a major step towards justice. Specifically, improvement in diagnosis will lead to reduction in discrimination, misclassification and prejudice experienced by patients, and better provide freedom from misdiagnosis to patients. 
\\ \\ 
     Augmented cognition, augmented gesture and augmented touch from neurotechnology will allow physicians to perform treatments and therapy in a more efficient and effective manner. The benefits of performing successful treatments and therapy are similar to those of performing a successful surgery or diagnosis. Just as in performing a successful surgery provides justice to the patient by allowing them to leave the hospital and get back to their routinely activities, performing a successful treatment is no different. However, given the difference in nature of surgeries and treatments in terms of activities and procedures, there are particular benefits that neurotechnology and augmentations provides that applies more for physicians than for surgeons. One of these benefits is how they help physicians with meeting their fiduciary duty to their patients during treatment through enabling physicians to properly implement precision medicine within the treatment itself. What makes precision medicine desirable for treatments is that its approach consists of utilizing the patient’s genes, environment, and lifestyle. This is especially of interest for patients as they would much rather have a treatment that is personalized for them than a one-size-fits-all treatment.  
\section{Ethical Analysis}
\subsection{Personhood Assessment}
    In the ethical evaluation of technological interventions, integrity and dignity of a person are the most relevant criteria [29]. Within philosophical literature, a portion of ethical questions raised in regard to neurotechnology are associated with what we call our “self” or “Soul”, which the debate itself usually draws on the concept of personhood as a modern notion that includes core aspects that we typically ascribe to our self or soul; these aspects include planning of the individual future and responsibility [29]. The concept of personhood always has normative implications due to the fact that we describe certain attributes and capabilities of a person, and want to have that person recognized, acknowledged and guaranteed [29]. Just as patients must consciously authorize a neurotechnological intervention before it is conducted, physicians and surgeons must also consciously provide authorization as well. As such, the concept of a person can provide an ethical benchmark applicable to persons such as physicians and surgeons, assuming that we do not want to impair the personal capabilities, such as autonomy and responsibility interventions in the brain [29]. With that being said, neurotechnological interventions are ethically unacceptable if remaining a person is at risk.
\\ \\
     This statement poses major problems for neurotechnology and human augmentation for physicians and surgeons for two reasons. To best understand those reasons, we will provide two cases. Note that the concept of personal identity refers to the query to which degree and under which circumstances a person remains the same over time, above and beyond physical identity  [29]. In investigating and evaluating for the “sameness” for a person, we need to consider both the interaction that the person has with others and the appreciation of moral capabilities, such as the ability to make a promise and keep it  [29]. This, especially, is important for the physicians and surgeons as both of them have a fiduciary responsibility to their patients. Consider the simple case where all physicians and surgeons provide authorization for augmentation via neurotechnology and neurotechnology successfully accomplishes its augmentations on them without any interference (i.e. no hackers or third parties involved). An undesired outcome that can be produced from this comes from the augmented cognition. The increase in knowledge does not guaranteed consistency or improvement in morality within a person. As such, there is a chance that the cognitive enhancement that they have undergone changes the way that they appreciate moral capabilities. A physician with increase in knowledge on the importance of using a specific biomedical approach to treating patients with a specific condition can reinforce their current belief that biomedical approaches are better than holistic approaches. In turn, this can lead to a lack of openness for alternative approaches such as a holistic approach, and empathy needed to understand why approaches like the holistic approach is needed for the treatment of those kinds of patients. This will lead to disputes and disagreements between the physician or surgeon and other physicians and surgeons, and with medical practitioners such as nurses and clinical social workers. Overtime, this will lead to the interaction that they have with each other as well as with those medical professionals to change. By the end of it, at the cost of wanting to improve their cognitive abilities for the sake of their patients, the physician or surgeon will have lost their personal identity. This possible loss of personal identity makes it problematic for neurotechnological interventions as the physician or surgeon is at risk of not remaining the person that he or she was prior to the augmentation.
\\ \\  
     For the second case, consider the instance where simple case where all physicians and surgeons provide authorization for augmentation via neurotechnology and neurotechnology successfully accomplishes its augmentations on them, but an interference by a third party, by say a malicious brain-hacker, occurs later. After such an instance occurs, assuming that the negotiation with the malicious brain-hacker was successful resulting in no changes made to the BCI and brain of the physician or surgeon, it is pretty much guaranteed that the physician or surgeon will not remain after the incident. The physician or surgeon will likely suffer from sort of paranoia, trust issues, and/or psychological trauma. This is not surprising as often when an individual’s mental integrity and dignity has been compromised, they end up suffering mentally from it to some degree. Furthermore, with the advancement of technologies and areas within computer science, specifically artificial intelligence, machine learning and quantum computing, it is justified for them to have fears and worries as there is no guarantee that the same issue will not be experienced again. What that degree is heavily depends on the person’s mental strength, willpower and self-control, the amount of support provided to them after the incident has occurred, and the environment where the person is situated after the occurrence of the incident. Regardless, their interaction with physicians, surgeons and others will likely change and their appreciation of moral capabilities will be altered as their view on them will be called into question. In worst case scenario, they become diagnosed with a psychological or neurological disorder as an aftermath or have their brain altered by the malicious brain-hacker, guaranteeing that their interaction with others and their appreciation of moral capabilities will change. This unintentional or intentional loss of personal identity makes it problematic for neurotechnological interventions as the physician or surgeon is at risk of becoming a new person that he or she was not prior to the augmentation. 
\subsection{Social Issues}
    What can be noticed in the previous section is a third reason, or case as we can call it, which is the case where some physicians and surgeons decides to either not provide authorization to the neurotechnological intervention altogether or to provide authorization to the neurotechnological intervention but not to undergo at least one type of augmentation. The types of groups and a possible distribution of these groups is shown in Figure 2. This case leads to one of the previously raised concerns stemming from the consequence of human augmentation. As discussed earlier, there is a chance that augmentations can lead to a divide between humans, where the population of humans with augmentations will feel that they are more superior to those without augmentation, leading to different types of social issues. In our case with the physician or surgeon not 
\begin{figure}
  \includegraphics[width=\linewidth]{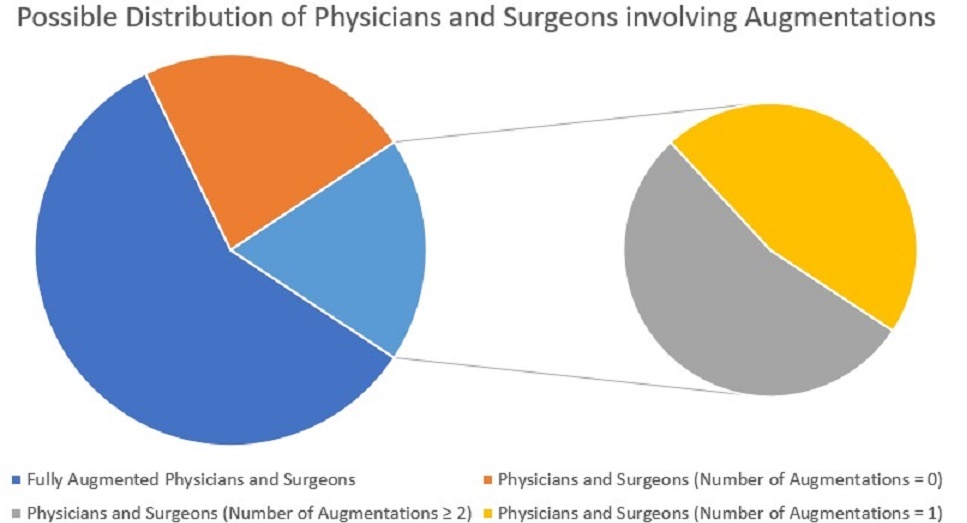}
  \caption{ Diagram displaying the types of groups that would exist in the case where undergo types of augmentations for physicians and surgeons are optional, and a possible distribution of each group. To avoid providing false information, population percentages were omitted. 
}
  \label{fig:figure2}
\end{figure}
 providing authorization to neurotechnological intervention, this divide can occur in three different divides: 
\begin{enumerate}
  \item Division among physicians and surgeons domestically. 
  \item Division between physicians and surgeons, and healthcare professionals. 
  \item Division between physicians, surgeons and healthcare professionals. 
  \item Division among physicians and surgeons internationally.  
\end{enumerate}
     For our case, we will only mainly focus on (1) and (4) as (2) and (3) are more distant to the scope of the paper. However, those are areas that can and should be further investigated. With there being four different groups, there is the change that those that physicians and surgeons that are fully augmented (i.e. have successfully undergone all augmentation via neurotechnology) mistreat those that do not have the augmentations that they have. This possibility is not farfetched and is very likely to happen. In every industry, including healthcare and medicine, we see prejudice and workers being mistreated due to their title, lack of experience or the lack of knowledge that (at least according to the administration or management within their organization). With those physicians and surgeons that do not have a particular augmentation, such as and especially augmented vision, cognition and touch, fully and semi augmented physicians and surgeons can take advantage of that to enforce their agendas and ideologies and bypassing arguments and statements made by them on the grounds that given their inferiors cognitive abilities or senses, what they say and their opinions should not be weighted equally to those that are fully augmented, or things along those lines. As such, augmentations increase the chance of prejudice and mistreatment occurring among physicians and surgeons, reinforcing existing beneficence, non-maleficence and justice issues within healthcare and medicine. Overtime, this can lead to alteration of personality among physicians and surgeons. Furthermore, this can lead to moral distress, trust issues, and other psychological issues experienced by those doctors that are mistreated, resulting in a loss of personal identity. This is dangerous as this shift in personality and personal identity will not only impact the performance of those physicians and surgeons and the lives of patients, but can create or further contribute to suicidal thoughts within them. Depending on the profile, involvements, commitments and types of patients that the doctor was treating, the impact that their suicide can have can be one that is negatively impacts individuals on a one-to-one, domestic, and/or international level. 
\\ \\ 
     The other way in which physicians or surgeons can be discriminated is in an instance where either malicious brain-hacking or a data hostage leading, with either of them leading to data sharing occurs. What makes doctors neural data valuable is that the doctors are known for being very well off financially. With proper preprocessing of the neural data and correct use of data mining and machine learning algorithms, health status, preferences and behavior of physicians and surgeons, physicians and surgeons can be exploited financially by companies such as insurance companies. Note that data records are stored electronically, which the highly personalized nature of neural data – much like genomic data – may increase the identifiability of individuals [25]. With that being said, the neural data obtained from malicious brain-hacking or data sharing could be used for discrimination purposes against a particular physician or surgeon, or to blackmail them into doing something. The severity of the consequences from the blackmail will depend on what is being asked and whether the doctor carries out the act. Regardless of what is produced from this scenario, as the risk of psychological harms depends directly on the presence of other risks [16], it is guaranteed that the doctor will experience distress, anxiety and other psychological issues, which can lead decreased performance in completing medical and non-medical tasks. This decrease in performance increases the chances that they harm the patient. 
\subsection{Global Health Inequality and Health Migration}
     Given that there are major economical, technological, and infrastructural disparities among countries, it is inevitable that some countries will not be able to provide their physicians and surgeons with the neurotechnological interventions needed to obtain augmentations. As computing technologies can greatly contribute to the growth and development of countries [4], countries such as the United States, Germany and China that have the manpower, infrastructure and resources to realize computing technologies for economic, social and medical development will be able to provide their doctors with the augmentations quicker and with guarantees compared to countries that are struggling due to poverty, warfare and humanitarian crises such as Libya, Syria and Myanmar. Consequently, if only doctors in developed countries are augmented, then this will create a global health inequality. Patients in countries with augmented physicians and surgeons will receive better treatment than those with regular physicians and surgeons. Furthermore, in times where doctors from different countries have to collaborate together on medical and healthcare issues, conflicts will likely arise between augmented doctors from one country (i.e. developed country) versus regular doctors from another (i.e. underdeveloped country). With augmented doctors possibly having a lack of openness for alternative approaches and empathy towards patients from different countries, this will lead to disputes and disagreements between the doctors from the different countries. The disputes and disagreements can lead to physicians and surgeons experiencing distress, anxiety and other psychological issues, resulting in decreased performance in completing medical and non-medical tasks. This decrease in performance increases the chances that they harm their patients and patients internationally depending on the case. 
\\ \\ 
     With developed and high-income countries being able to provide not just augmentations to physicians and surgeons but also better employment opportunities, education, safety and security for them and their families, these will push physicians and surgeons to migrate to those countries [34]. Their migration will further contribute to the current disparity in the health workforce between high- and low/middle-income countries [35]. In Serour’s work, which was done back in 2009, she stated that Africa needs 1 million health workers, which includes physicians and surgeons, for Sub-Saharan Africa alone. This number has likely increased due to the SARS-CoV-2 outbreak that has and continues to take place in 2020. With countries like the United States, United Kingdom, Canada and Australia already benefitting considerably from the migration of nurses and doctors [30], the implementation of neurotechnology for human augmentation will further benefit those countries as well as other developed countries, leaving the underdeveloped countries to suffer more. What results from this is an increase in harm experienced by the patients and an increase in mortalities in the underdeveloped countries’ population.
\subsection{Patient Harm}
    Whether it is data hostage situation of the data from a TMS or malicious brain-hacking of a neuroprosthetic, given the risks associated with physicians and surgeons using neurotechnology for augmentations along with the impact that augmentation has on people in general, there is no doubt that there is a chance that the change in self-concept, personal identity or the discrimination, psychological and other issues experienced by physicians and surgeons will influence doctor-patient relationships. One of the reasons for why the relationship will likely become worse is because the voices of the patients will either be ignored or be of less value to their physicians and surgeons. For fully or semi augmented physicians and surgeons, especially those with augmented cognition, there is a higher chance that they will think that whatever decision they make will be the right decision for the patient. Consequently, when patients of theirs voice their concern to them, they will either ignore or listen only selected parts of the concern. In turn, this will make the patient feel as if their doctor does not care for them, leading to feelings of sadness, distress, and pain. This is problematic because even in the case where the patient is successfully treated, they would leave the hospital or clinic feeling mentally hurt. While the degree in which the mental pain from the experienced differs from one patient to the other, it is still a problem because the physician or surgeon did not accomplish non-maleficence. During times when they could have prevented the patient from experiencing mental harm, they did not prevent it. As a result, the physician or surgeon did not ensure to inflict the least amount of harm possible to reach a beneficial outcome. Rather, they inflicted an amount of harm onto the patient that either they were not aware of or did not care about enough to reach a beneficial outcome.
\\ \\ 
     The other reason for why doctor-patient relationships will likely worsen is because of the side effects experienced by physicians and surgeons from either undergoing all augmentations or selected few augmentations, and/or from being mistreated and discriminated by their colleagues [15]. For fully or semi augmented physicians and surgeons, in  the case where they lose their personal identity, this can make it difficult for patients to communicate their concerns as the new personality developed by their physician or surgeon can be one that makes it difficult to communicate with them, and vice-versa. Consequently, in the  physician or surgeon having difficulty communicating with their patients, this increases the chances of  and eventually lead them harm their patients either by accident or on purpose. In the case of harm occurring due to an accident, this accident could be due to the physician or surgeon making a wrong assumption based on the limited information verbally provided by their patients. Alternatively, this can also be due to after using tES to increase the physician or surgeon memory, rather than the memory being used for remembering details about the patients, the memory retrieval within the physician or surgeon’s brain retrieves unrelated information or even problematic information such as details of a traumatic event; in retrieving the wrong information, the physician or surgeon’s attention is not adequately concentrated on the patient, making it difficult for them to communicate with the patient, and eventually leading to them accidentally harming the patient. In the case of harm occurring on purpose, this could be due to first the personal identity change resulting in the physician or surgeon not being able to properly control their anger, then the patient criticizing the physician or surgeon, infuriating the doctor in a way that made them feel that the patient is attacking their expertise, experiences and augmentations. 
\\ \\
     The problem of the voices of patients being ignored extends to issues of justice. What results from the patient’s voice being partially or fully ignored is unequal and unfair treatment by the fully or semi augmented physicians and surgeons to their patients. The way that they approached the medical intervention was not the same across all of their patients. Some received more care from the physicians or surgeons while others received less. Consequently, this unequal level of care is unfair to those that do not receive the same amount of care from the physicians or surgeons. This unequal treatment to patients can not only be experienced while undergoing the medical intervention, but it can be experienced prior to it as well. Especially with augmented cognition, augmented vision and BCIs, those make it possible for physicians and surgeons to easily take advantage of their patients during the diagnosis and consent phase prior to the treatment. With tDCS providing physicians and surgeons the ability to better solve complex problems, they can find ways to trick patients into providing their “informed” consent to treatments that are either costly, ineffective, harmful or all three combined. Consequently, this will lead to the patient further suffering from the disease and any complications that arise during the medical intervention, financial loss, and increased levels of distress from spending a larger than expected time in the hospital and from the worry of either dying or possibly not being able to live a regular life again or anytime soon. 
\section{Conclusion}
    A limitation with our work is that there were no medical professionals that directly contributed to the paper. As such, there are some notable ethical issues that were not included in this paper. Additionally, as our work did not include any discussions on health policy and cyber security, we were not able to discuss solutions that could be used to address the ethical and health issues. Future work can ethically analyze the impact of neurotechnology for human augmentation on physicians and surgeons with respect to the division between physicians, surgeons and healthcare professionals, and should focus on health policy and cyber security needed to address those ethical and health issues. Even with the best-case scenario where neurotechnology always succeeds in providing people with augmentations, in applying it on physicians and surgeons to augment them, there are high risk associated with it. This makes it challenging when it comes to doing a cost benefit analysis of it, especially because when doing such an analysis, we cannot make the big assumption that we made in this paper. A way to reduce the risks associated with the neuroengineering component of augmenting physicians and surgeons is to design and program the tactile sensors and neuroprosthetics so that in a worse case where a malicious brain-hacker takes over a surgeon’s BCI and tries to make the surgeon harm their patient, the neuroprosthetic detects the abnormality then communicates with the tactile sensors, activating the emergency magnet mechanism in the glove, forcing the gloves to tighten together like handcuffs. This will prevent the surgeon from harming the patient and medical practitioners in the room, and give the surgical team time to figure out what to do next. It is important that such safety mechanisms like these and others are rigorously tested prior to using them in clinical settings, and have them ready for use once they are ready to be used. Furthermore, after augmentations have been embedded within physicians and surgeons, have a designated medical and bioethics combined team dedicated to supporting them mentally to ensure that the transition is smooth and the physicians and surgeons’ personal identity and other crucial aspects of them are preserved. In doing so, this can help reduce and prevent the occurences of harm and ethical, medical, and personhood issues experienced by physicians, surgeons and patients.
%
%

%
\end{document}